\documentclass[11pt]{article}%
\usepackage{amsmath}
\usepackage{amsfonts}
\usepackage{amssymb}
\usepackage{amsthm}
\usepackage{graphicx}%
\providecommand{\U}[1]{\protect\rule{.1in}{.1in}}
\providecommand{\U}[1]{\protect\rule{.1in}{.1in}}
\newtheorem{theorem}{Theorem}[section]

\theoremstyle{definition}

\theoremstyle{remark}

\newtheorem{remark}[theorem]{Remark}
\numberwithin{equation}{section}
\begin{document}

\title{Analysis of graded-index optical fibers by the spectral parameter power series method}
\author{Ra\'{u}l Castillo-P\'{e}rez$^{1}$, Vladislav V. Kravchenko$^{2}$ and Sergii M.
Torba$^{2}$\\{\small $^{1}$SEPI, ESIME Zacatenco, Instituto Polit\'{e}cnico Nacional, Av.
IPN S/N, C.P. 07738, D.F. Mexico}\\{\small $^{2}$Departamento de Matem\'{a}ticas, CINVESTAV del IPN, Unidad
Quer\'{e}taro, Libramiento Norponiente \#2000,}\\{\small Fracc. Real de Juriquilla, Quer\'{e}taro, Qro. C.P. 76230 MEXICO}\\e-mail: rcastillo@ipn.mx, vkravchenko@math.cinvestav.edu.mx,\\storba@math.cinvestav.edu.mx}
\maketitle

\begin{abstract}
Spectral parameter power series (SPPS) method is a recently introduced
technique \cite{KrCV08}, \cite{KrPorter2010} for solving linear differential
equations and related spectral problems. In the present work we develop an
approach based on the SPPS for analysis of graded-index optical fibers. The
characteristic equation of the eigenvalue problem for calculation of guided
modes is obtained in an analytical form in terms of SPPS. Truncation of the
series and consideration in this way of the approximate characteristic
equation gives us a simple and efficient numerical method for solving the
problem. Comparison with the results obtained by other available techniques
reveals clear advantages of the SPPS approach, in particular, with regards to
accuracy. Based on the solution of the eigenvalue problem, parameters
describing the dispersion are analyzed as well.

\end{abstract}

\section{Introduction}

Analysis of graded-index cylindrical waveguides, in particular of optical
fibers, presents considerable mathematical and computational difficulties. The
main differential equation involved is singular, the fact which restricts
application of purely numerical techniques. The typically used approach is
based on the asymptotic WKB approximation (Wentzel--Kramers--Brillouin). It
can offer analytical relations between the mode-propagation parameters (see,
e.g., \cite{Marcuse}, \cite[Sect. 3.7]{Okamoto}). However, it is well known
(see, e.g., the discussion in \cite{Liu et al}) that they are not accurate
enough for the practical study of optical fibers.

We develop a completely new approach for analysis of the computationally
difficult spectral problems involved. It is based on the spectral parameter
power series (SPPS) representation for solutions of second-order\ linear
differential equations of the Sturm-Liouville type. The SPPS representation
for regular Sturm-Liouville equations was first obtained in \cite{KrCV08} and
applied to numerical study of spectral problems in \cite{KrPorter2010}. It was
used in a number of applications (we refer to the review \cite{KKR2012}). In particular in \cite{CKKO2009} the SPPS method was applied to the study of wave propagation in layered media. In
\cite{CKT2013} the SPPS representation was obtained for solutions of the
singular differential equations belonging to the class of perturbed Bessel
equations. The main differential equation of the graded-index cylindrical
waveguides is of that type. In the present paper we use the main result of
\cite{CKT2013} to develop a numerical method for analysis of graded-index
optical fibers. The characteristic equation equivalent to the spectral problem
for calculation of guided modes is obtained in an explicit analytical form.
The numerical method consists in approximating this equation and finding zeros
of the approximate characteristic function.

The paper contains several numerical tests. In the case of the well studied
parabolic profile we compare our results with the available exact solution,
with the WKB approach as well as with the results from \cite{Liu et al} where
the finite element method was applied with the results of the WKB
approximation used as the initial guess. Our method gives substantially more
accurate values.

We show that the SPPS method presented here allows one a quick and accurate
analysis of graded-index optical fibers and can be used as an efficient tool
for their design.

\section{Mathematical formulation of the problem}

The basic wave equation governing the wave propagation in graded-index fibers
has the form (see, e.g., \cite[Sect. 3.7]{Okamoto})%
\begin{equation}
\frac{d^{2}}{dr^{2}}\psi+\frac{1}{r}\frac{d}{dr}\psi+\left(  k^{2}%
n^{2}(r)-\beta^{2}-\frac{m^{2}}{r^{2}}\right)  \psi=0,\qquad r\in(0,a],
\label{main wave equation}%
\end{equation}
where the knowledge of the solution $\psi$ allows one to compute the
corresponding electromagnetic field, $n(r)$ is the radial refractive index
profile, $k=2\pi/\lambda$ is the vacuum wave number, $\beta$ is the
propagation constant, and $m$ is a mode parameter given by%
\[
m=\left\{
\begin{array}
[c]{ll}%
1 & \text{TE and TM modes (}\ell=0\text{)}\\
\ell+1 & \text{EH mode (}\ell\geq1\text{)}\\
\ell-1 & \text{HE mode (}\ell\geq1\text{)}.%
\end{array}
\right.
\]
As in most practical fibers the refractive index varies in the core but is
constant in the cladding, it is usually convenient to solve the wave equation
in the core and cladding separately and to match those solutions at the
core-cladding boundary. If the solutions in the core and cladding are denoted
by $\psi(r)$ and $\psi_{\text{clad}}(r)$, respectively, the boundary
conditions at the interface $r=a$ under the weakly guiding approximation (see, e.g., \cite[p.
94]{Okamoto}) are given by%
\begin{equation}
\psi(a)=\psi_{\text{clad}}(a)\text{\quad and\quad}\left.  \frac{d\psi}%
{dr}\right\vert _{r=a}=\left.  \frac{d\psi_{\text{clad}}}{dr}\right\vert
_{r=a}. \label{boundary cond 1}%
\end{equation}
We note that for the method presented here it is not essential that the
boundary conditions be that simple. For example, they could be dependent on
$k$.

Introducing the function $U(r)=\sqrt{r}\psi(r)$ we write equation
(\ref{main wave equation}) in the form%
\[
-U^{\prime\prime}+\left(  \beta^{2}+\frac{m^{2}-1/4}{r^{2}}-k^{2}%
n^{2}(r)\right)  U=0.
\]
Finally, introducing the new independent variable $x=r/a$ we obtain an
equation for the function $u(x)=U(ax)$,%
\begin{equation}
-u^{\prime\prime}+\left(  a^{2}\beta^{2}+\frac{m^{2}-1/4}{x^{2}}-a^{2}%
k^{2}n^{2}(ax)\right)  u=0. \label{main perturbed Bessel equation}%
\end{equation}
The boundary conditions (\ref{boundary cond 1}) take the form%
\begin{equation}
u(1)=u_{\text{clad}}(1)\text{\quad and\quad}\left.  \frac{du}{dx}\right\vert
_{x=1}=\left.  \frac{du_{\text{clad}}}{dx}\right\vert _{x=1}.
\label{boundary cond}%
\end{equation}
The solution for the cladding ($x>1$) has the form%
\[
u_{\text{clad}}(x)=C\sqrt{x}K_{m}\Bigl(a\sqrt{\beta^{2}-k^{2}n_{2}^{2}%
}\,x\Bigr)
\]
where $K_{m}$ is the modified Bessel function of the second kind, the constant
$n_{2}$ is the value of the refractive index in the cladding, and $C$ is an
arbitrary constant.

The bounded solution $u$ in the core is unique up to a multiplicative
constant. Thus, from the boundary conditions (\ref{boundary cond}) the
following characteristic equation of the problem is obtained%
\[
\frac{u^{\prime}(1)}{u(1)}=\frac{u_{\text{clad}}^{\prime}(1)}{u_{\text{clad}%
}(1)},
\]
which then admits the form
\begin{multline}
2K_{m}\Bigl(a\sqrt{\beta^{2}-k^{2}n_{2}^{2}}\Bigr)u^{\prime}(1)-\\
\left(  \left(  1+2m\right)  K_{m}\Bigl(a\sqrt{\beta^{2}-k^{2}n_{2}^{2}%
}\Bigr)-2a\sqrt{\beta^{2}-k^{2}n_{2}^{2}}K_{m+1}\Bigl(a\sqrt{\beta^{2}%
-k^{2}n_{2}^{2}}\Bigr)\right)  u(1)=0. \label{characteristic equation}%
\end{multline}
Pairs of values of $\beta$ and $k$ which satisfy this characteristic equation
together with the light propagation condition
\begin{equation}
k^{2}n_{2}^{2}<\beta^{2}<k^{2}n_{1}^{2}:=k^{2} \max_{0\le x\le1} n^{2}(ax)
\label{condition for Beta}%
\end{equation}
correspond to guided modes in the fiber and are the main object of
computation. As we show below, the SPPS approach allows one to approximate
directly the characteristic equation and to solve it with a considerable
accuracy and speed.

\begin{remark}
Note that the characteristic equation \eqref{characteristic equation} remains
valid also for the case of absorbing media, i.e., when the refractive index
$n(r)$ has non-zero imaginary part. The only change is in the propagation
condition \eqref{condition for Beta}, it should be written as
\[
\operatorname{Re} n_{2}^{2}<\frac{\beta^{2}}{k^{2}}<\max_{0\le x\le1}
\operatorname{Re} n^{2}(ax).
\]

\end{remark}

\section{Analysis of dispersion}

The solution of the characteristic equation (\ref{characteristic equation})
gives us the dependence of $k$ on $\beta$ (or, since $k=2\pi/\lambda=\omega
/c$, the dependencies of $\lambda$ and $\omega$ on $\beta$). A way to employ
the obtained information is in the analysis of dispersion. The modal
description of dispersion is related to the different mode indices (or group
velocities) associated with different modes. The group velocity associated
with the fundamental mode is frequency dependent because of chromatic
dispersion. As a result, different spectral components of the pulse travel at
slightly different group velocities, a phenomenon referred to as group
velocity dispersion (GVD) which has two contributions: material dispersion and
waveguide dispersion \cite{Agrawal}. Group velocity $v_{g}$ is defined as
\cite{Born}%
\begin{equation}
v_{g}=d\omega/d\beta\label{Group Velocity}%
\end{equation}
where $\omega$ stands for the frequency. If $\Delta\omega$ is the spectral
width of the pulse, the extent of pulse broadening for a fiber of length $L$
is governed by \cite{Agrawal}%
\begin{equation}
\Delta T=\frac{d}{d\omega}\left(  \frac{L}{v_{g}}\right)  \Delta\omega
=L\frac{d^{2}\beta}{d\omega^{2}}\Delta\omega=:L\beta_{2}\Delta\omega.
\label{DeltaT}%
\end{equation}
where $\Delta T$ stands for the differential group delay and the parameter
$\beta_{2}=d^{2}\beta/d\omega^{2}$ is known as the GVD parameter. In terms of
the range of wavelengths $\Delta\lambda$ emitted by the optical source, and by
using $\omega=2\pi c/\lambda$ and $\Delta\omega=(-2\pi c/\lambda^{2}%
)\Delta\lambda$, (\ref{DeltaT}) can be written as%
\[
\Delta T=\frac{d}{d\lambda}\left(  \frac{L}{v_{g}}\right)  \Delta
\lambda=DL\Delta\lambda,
\]
where the dispersion parameter $D,$ expressed in units of ps/(km-nm), is given
by%
\begin{equation}
D=\frac{d}{d\lambda}\left(  \frac{1}{v_{g}}\right)  =-
\frac{2\pi c}{\lambda^{2}}\beta_{2}=
-\frac{\lambda}{2\pi c}\left(2\frac{d\beta}{d\lambda}+\lambda\frac{d^2\beta}{d\lambda^2}\right). \label{DispersionParameter}
\end{equation}
Interest in studying fibers with different refractive index profiles comes
from the fact that tayloring such profiles the dispersion, which is one of the
main impairments for optical communications, can be manipulated. Fibers where
the dispersion parameter can be reduced at a certain wavelenght, flattened for
a wide range of wavelenghts or even made highly negative are used for
aplications concerning long haul communications, wavelenght division
multiplexing and dispersion compensation, among others (see, e.g.,
\cite{Ainslie Day},\cite[Chapter 4]{Oh and Paek}).

\section{Preliminary facts and SPPS representations for perturbed Bessel
equations}

\label{SectSPPS}

The main equation (\ref{main perturbed Bessel equation}) is of the form
\begin{equation}
-u^{\prime\prime}+\left(  \frac{m^{2}-1/4}{x^{2}}+q(x)\right)  u=\mu
r(x)u,\qquad x\in(0,a], \label{perturbed Bessel}%
\end{equation}
where $m\in\mathbb{N}$, $q$ and $r$ are known functions (which in general can
be complex valued), $\mu$ is a spectral parameter. The equation belongs to the
class of perturbed Bessel equations and was studied in a considerable number
of publications (e.g., \cite{BoumenirChanane}, \cite{CKT2013},
\cite{Chebli1994}, \cite{Guillot 1988}, \cite{KosTesh2011}, \cite{Weidmann}).
In what follows we always assume that $q$ and $r$ are piecewise continuous
functions on the whole segment of interest with at most a finite number of
step discontinuities.

The SPPS method applied to (\ref{perturbed Bessel}) consists of two steps. On
the first step one needs to construct a nonvanishing on $(0,a]$ bounded
solution $u_{0}$ of the equation
\begin{equation}
-u_{0}^{\prime\prime}+\left(  \frac{m^{2}-1/4}{x^{2}}+q(x)\right)  u_{0}=0.
\label{equation for particular solution}%
\end{equation}
In \cite{CKT2013} it was proved that if $q(x)\geq0$, $x\in(0,a]$ then such
solution $u_{0}$ exists and has the form
\begin{equation}
u_{0}(x)=x^{m+1/2}\sum_{k=0}^{\infty}\widetilde{Y}^{(2k)}(x),
\label{FnPartSol}%
\end{equation}
where the functions $\widetilde{Y}^{(j)}$ are defined recursively as follows
\begin{equation}%
\begin{split}
\widetilde{Y}^{(0)}  &  \equiv1,\\
\widetilde{Y}^{(j)}(x)  &  =%
\begin{cases}
\displaystyle\int_{0}^{x}\widetilde{Y}^{(j-1)}(t)t^{2m+1}q(t)\,dt, & \text{for
odd }j,\\
\displaystyle\int_{0}^{x}\widetilde{Y}^{(j-1)}(t)t^{-(2m+1)}\,dt, & \text{for
even }j.
\end{cases}
\end{split}
\label{YtildePS}%
\end{equation}
The series (\ref{FnPartSol}) converges uniformly on $[0,a]$. When $q$ does not
satisfy the nonnegativity condition the series (\ref{FnPartSol}) still defines
a bounded solution of (\ref{equation for particular solution}) but in general
not necessarily nonvanishing on $(0,a]$. The derivative of $u_{0}$ has the
form%
\begin{equation}
u_{0}^{\prime}(x)=\left(  m+\frac{1}{2}\right)  x^{m-1/2}\sum_{k=0}^{\infty
}\widetilde{Y}^{(2k)}(x)+x^{-(m+1/2)}\sum_{k=1}^{\infty}\widetilde{Y}%
^{(2k-1)}(x). \label{der u0}%
\end{equation}

On the second step the solution $u_{0}$ is used for constructing the unique
(up to a multiplicative constant) bounded solution of (\ref{perturbed Bessel})
for any value of $\mu$. This solution has the form
\begin{equation}
u(x)=u_{0}(x)\sum_{k=0}^{\infty}\mu^{k}\widetilde{X}^{(2k)}(x) \label{SPPSsol}%
\end{equation}
where the functions $\widetilde{X}^{(j)}$ are defined recursively as follows
\begin{equation}%
\begin{split}
\widetilde{X}^{(0)}  &  \equiv1,\\
\widetilde{X}^{(j)}(x)  &  =%
\begin{cases}
\displaystyle\int_{0}^{x}u_{0}^{2}(t)r(t)\widetilde{X}^{(j-1)}(t)\,dt, &
\text{if }j\text{ is odd},\\
-\displaystyle\int_{0}^{x}\frac{\widetilde{X}^{(j-1)}(t)}{u_{0}^{2}(t)}\,dt, &
\text{if }j\text{ is even}.
\end{cases}
\end{split}
\label{Xtilde}%
\end{equation}
The series (\ref{SPPSsol}) converges uniformly on $[0,a]$. The first
derivative of $u$ is given by
\begin{equation}
u^{\prime}=\frac{u_{0}^{\prime}}{u_{0}}u-\frac{1}{u_{0}}\sum_{k=1}^{\infty}%
\mu^{k}\widetilde{X}^{(2k-1)}, \label{SPPSsolDer}%
\end{equation}
where the series converges uniformly on an arbitrary compact $K\subset(0,a]$.

In (\ref{SPPSsol}) the solution is represented in the form of a power series
with respect to the spectral parameter $\mu$. Such representations were called
in \cite{KrPorter2010} the spectral parameter power series (SPPS). The
representation (\ref{SPPSsol}) is based on a particular solution of equation
(\ref{perturbed Bessel}) for $\mu=0$. Similarly to the Taylor series, the
approximations given by the truncation of the series \eqref{SPPSsol} and
\eqref{SPPSsolDer} are more accurate near the origin, while the accuracy
deteriorates with the increase of the parameter $\mu$, see \cite[Example
7.7]{CKT2013}. In \cite{CKT2013, KrPorter2010} it was mentioned that it is
also possible to construct the SPPS representation of a general solution
starting from a non-vanishing particular solution for some $\mu=\mu_{0}$. Such
a procedure is called a spectral shift and consists in the following. Equation
(\ref{perturbed Bessel}) can be written in the form
\begin{equation}
-u^{\prime\prime}+\left(  \frac{m^{2}-1/4}{x^{2}}+q(x)-\mu_{0}r(x)\right)
u=\left(  \mu-\mu_{0}\right)  r(x)u \label{eq Bessel shifted}%
\end{equation}
where $\mu_{0}$ is an arbitrary number. This equation is again of the form
(\ref{perturbed Bessel}) but with the spectral parameter $\Lambda:=\mu-\mu
_{0}$. The SPPS technique applied to this equation leads to a representation
of the solution in the form of a power series in terms of $\Lambda$ and allows one
to improve the accuracy of the approximations given by the truncated series
\eqref{SPPSsol} and \eqref{SPPSsolDer} near the point $\mu_{0}$. The spectral
shift has already proven its usefulness for numerical applications
\cite{KrPorter2010}, \cite{KKB2013}, \cite{CKT2013} and is used below in
numerical computations.

\begin{remark}
\label{Non vanishing PS} Since equation \eqref{perturbed Bessel} for each
$\mu$ possesses a unique (up to a multiplicative constant) bounded solution,
it might look difficult to find a $\mu_{0}$ providing a non-vanishing
particular solution. The sufficient condition $q(x)-\mu_{0}\ge0$ mentioned
above imposes an upper bound for the possible choices of the parameter
$\mu_{0}$. Allowing $\mu_{0}$ to be complex valued can solve the problem.
Assume that $q$ is real valued and $r(x)>0$ almost everywhere on $(0,a]$ (which is
sufficient for the scope of the paper). Then every particular bounded solution
of equation \eqref{perturbed Bessel} having $\mu$ such that $\operatorname{Im}%
\mu\ne0$ is non-vanishing on $(0,a]$. Indeed, the equality $u(x_{0})=0$,
$x_{0}\in(0,a]$ contradicts the well-known property (see, e.g., \cite[Chapter
10.4]{Zettl}) of the operator $\tau u = -u^{\prime\prime}+\left(  \frac
{m^{2}-1/4}{x^{2}}+q(x)\right)  u$ with a suitable domain to be self-adjoint
in the Hilbert space $L_{2}((0,x_{0}], r)$ and have only real eigenvalues.
\end{remark}

\section{The SPPS form of the characteristic equation}

Let the parameter $\beta$ be fixed. Writing equation
\eqref{main perturbed Bessel equation} as
\[
-u^{\prime\prime}+\left(  \frac{m^{2}-1/4}{x^{2}}+a^{2} \beta^{2}\right)  u =
a^{2}k^{2}n^{2}(ax)u
\]
we obtain an equation in the form considered in Section \ref{SectSPPS} with
the spectral parameter $\mu= a^{2}k^{2}$ and having known particular
non-vanishing solution $u_{0}=c \sqrt{x}I_{m}(a\beta x)$. However the direct
application of the SPPS method for the obtained equation leads to certain
difficulties. First, for the practically used fibers the value $a\beta$ is
quite large (in the examples below it is within the range $20-300$) leading to a rapidly growing particular solution $u_{0}$. Second, since we are looking for the values of the parameter $k$ satisfying \eqref{condition for Beta}, we cannot take advantage of the property of the SPPS representation to be the most accurate near the origin. To overcome these difficulties we apply the spectral shift technique.

Let us assume that the refractive index $n$ is real valued. By $n_{1}$ we
denote the maximum of $n$ in the core. Note that $n_{1}>n_{2}$. The main
equation (\ref{main perturbed Bessel equation}) can be written in the
following form%
\begin{equation}
-u^{\prime\prime}+\left(  \frac{m^{2}-1/4}{x^{2}}+a^{2}\left(  \beta^{2}%
-k_{0}^{2}n^{2}(ax)\right)  \right)  u=a^{2}\left(  k^{2}-k_{0}^{2}\right)
n^{2}(ax)u \label{main perturbed transformed}%
\end{equation}
where $k_{0}=\beta/n_{1}$. Then $q(x):=a^{2}\left(  \beta^{2}-k_{0}^{2}%
n^{2}(ax)\right)  \geq0$, $x\in(0,1]$. Introducing the notations $\mu
:=a^{2}\left(  k^{2}-k_{0}^{2}\right)  $ and $r(x):=n^{2}(ax)$ we find that
equation (\ref{main perturbed transformed}) is an equation of the form
(\ref{perturbed Bessel}) with $q$ satisfying the nonnegativity condition from
the preceding section. Hence the solution $u_{0}$ constructed as explained
above does not have other zeros on $[0,1]$ except at $x=0$ and the bounded
solution of (\ref{main perturbed transformed}) has the form (\ref{SPPSsol}).

Consequently, the characteristic equation (\ref{characteristic equation}) can
be written as follows
\begin{multline}
2K_{m}\Bigl(a\sqrt{\beta^{2}-k^{2}n_{2}^{2}}\Bigr)\left(  u_{0}^{\prime
}(1)\sum_{k=0}^{\infty}\mu^{k}\widetilde{X}^{(2k)}(1)-\frac{1}{u_{0}(1)}%
\sum_{k=1}^{\infty}\mu^{k}\widetilde{X}^{(2k-1)}(1)\right) \\
-\left(  \left(  1+2m\right)  K_{m}\Bigl(a\sqrt{\beta^{2}-k^{2}n_{2}^{2}}
\Bigr)-2a\sqrt{\beta^{2}-k^{2}n_{2}^{2}}K_{m+1}\Bigl(a\sqrt{\beta^{2}%
-k^{2}n_{2}^{2}}\Bigr)\right)  \times\\
u_{0}(1)\sum_{k=0}^{\infty}\mu^{k}\widetilde{X}^{(2k)}(1)=0.
\label{characteristic equation SPPS}%
\end{multline}

\begin{remark}
\label{Remark another transformation} Solution of equation
\eqref{characteristic equation SPPS} gives values of the parameter $k$ for a
fixed $\beta$. For some applications it can be more convenient to find the
values of the parameter $\beta$ satisfying characteristic equation
\eqref{characteristic equation} for a fixed $k$. For this, equation
\eqref{main perturbed Bessel equation} can be written in the form
\[
-u^{\prime\prime}+\left(  \frac{m^{2}-1/4}{x^{2}}+a^{2}\left(  \beta_{0}^{2}
-k^{2}n^{2}(ax)\right)  \right)  u=a^{2}\left(  \beta_{0}^{2}-\beta
^{2}\right)  u,
\]
where $\beta_{0}^{2} = k^{2}n_{1}^{2}$. Then $q(x):=a^{2}(\beta_{0}^{2}
-k^{2}n^{2}(ax))\ge0$, $x\in(0,1]$ and the procedure described above
can be applied.
\end{remark}

\section{Numerical implementation and examples\label{SectNumeric}}

Based on the results of the previous sections we can formulate a numerical
method for computing the guided modes of the fiber. We suppose that some range
for the parameter $\beta$ is given, say $\beta\in[\beta_{1},\beta_{2}]$. If
instead a certain segment of wavelengths $[\lambda_{1}, \lambda_{2}]$ is
given, one can find the range of $\beta$'s from \eqref{condition for Beta}:
\begin{equation}
\label{Range for Beta}\beta\in\left[  \frac{2\pi n_{2}}{\lambda_{2}}%
,\frac{2\pi n_{1}}{\lambda_{1}}\right]  .
\end{equation}

\begin{enumerate}
\item For each $\beta$ belonging to a mesh over $[\beta_{1},\beta_{2}]$
perform steps \ref{Start m}--\ref{End SPPS}

\item Start with $m=0$ and perform steps \ref{Start SPPS}--\ref{End SPPS}
until no new propagation mode be found.\label{Start m}

\item Compute a particular solution $u_{0}$ of equation
(\ref{equation for particular solution}) according to (\ref{FnPartSol}) as
well as its derivative according to (\ref{der u0}).\label{Start SPPS}

\item Use partial sums of the series (\ref{SPPSsol}) and (\ref{SPPSsolDer}) to
obtain an approximation of the function appearing on the left-hand side of the
characteristic equation (\ref{characteristic equation SPPS}).

\item Find its zeros satisfying propagation condition
\begin{equation}
\label{Range for k}\frac{\beta^{2}}{n_{1}^{2}}<k^{2}<\frac{\beta^{2}}%
{n_{2}^{2}}.
\end{equation}

\item If the interval for the spectral parameter $\mu$ in
\eqref{main perturbed transformed} defined by \eqref{Range for k} is large
(e.g., $\max\mu>400$), perform several steps of the spectral shift technique
to improve the precision. A non-vanishing particular solution on each step can
be obtained using Remark \ref{Non vanishing PS}. \label{End SPPS}

\item Combine the values of parameter $k$ found for different values of $\beta$ to
obtain dependencies $k(\beta)$ for different propagation modes.
\end{enumerate}

Before considering numerical examples let us explain how the numerical
implementation of the SPPS method was realized  in this work. All calculations
were performed with the help of Matlab 2013b in the double precision machine
arithmetics in a PC with Intel i7-3770 processor. The formal powers
$\widetilde{X}^{(j)}$ were calculated using the Newton-Cottes 6 point
integration formula of 7$^{\text{th}}$ order, see, e.g.,
\cite{DavisRabinovich}, modified to perform indefinite integration. We choose
$M$ equally spaced points covering the segment of interest and apply the
integration formula to overlapping groups of six points. It is worth
mentioning that for large values of the parameter $m$ a special care should be
taken near the point $0$, because even small errors in the values of
$\widetilde{X}^{(2j-1)}$ after the division by $u_{0}^{2}\sim x^{2m+1}$ lead
to large errors in the computation of $\widetilde{X}^{(2j)}$ on the whole
interval $[0,1]$. To overcome this difficulty we change the values of
$\widetilde{X}^{(2j-1)}$ in several points near zero to their asymptotic
values. This strategy leads to a good accuracy. The computation of the first
100 to 200 formal powers proved to be a completely feasible task, and even for
$M$ being as large as several millions the computation time of the whole set
of formal powers is within seconds. In the presented numerical results we
specify, among others, two parameters: $N$ is the number of computed formal
powers $\widetilde{X}^{(j)}$, and $M$ is the number of points taken on the
considered segment for the calculation of integrals. Using the calculated
formal powers we evaluated the approximation of the characteristic equation
(\ref{characteristic equation SPPS}), constructed a spline passing throw the
obtained values and found its zeros using the Matlab command \texttt{fnzeros}.

\subsection{Mode characterization of a graded index fiber}

\label{Example 1}
As a first case of study, we present the computation of the propagation
constants $\beta$ for a graded index multimode fiber with a refractive index
given by%
\begin{equation}
n(x)=%
\begin{cases}
n_{1}\left(  1-2\Delta_{n}x^{\alpha}\right)  ^{1/2}, & 0\leq x\leq1\\
n_{2} & x>1
\end{cases}
\label{PowerProfiles}%
\end{equation}
where the relative refractive-index difference between $n_{1}$ and $n_{2}$ is
defined as%
\[
\Delta_{n}=\frac{n_{1}^{2}-n_{2}^{2}}{2n_{1}^{2}}.
\]
For $\alpha=2$ the exact solution of the problem is known allowing us to verify
the accuracy of the proposed method. The specific values that were used in
order to be able to compare our results with the ones in the literature and
with an exact solution were $n_{1}=1.462$, $n_{2}=1.447$, $a=25\,\mu$m, and
$\lambda=0.78\,\mu$m. The parameters associated to the numerical
implementation of the SPPS method included the approximation of the functions
with $M=100001$ points and using $N=100$ formal powers $\widetilde X^{(j)}$
for the truncated series. For this example it was natural to apply Remark
\ref{Remark another transformation}. Since the range of the new spectral
parameter $\mu= a^{2}(\beta_{0}^{2}-\beta^{2})$ defined by
(\ref{condition for Beta}) is approximately $[0, 1700]$, i.e., quite large, we
applied 8 steps of the spectral shift. As a result, on each step the range of
the spectral parameter was less than 200. The non-vanishing solutions were
obtained accordingly to Remark \ref{Non vanishing PS}, we used spectral shifts
$\mu_{n} = (200+4i)n$. On each step a 1000 points mesh was used to approximate
the characteristic equation.

\begin{table}[ptb]
\centering
\begin{tabular}
[c]{cllll}\hline
Mode & WKB & FEM & Exact & SPPS\\\hline
1 & 1.1770700 & 1.177085 & 1.177122819807467 & 1.177122819807467\\
2 & 1.1764113 & 1.176449 & 1.176550885053707 & 1.176550885053707\\
3 & 1.1757390 & 1.175795 & 1.175978672140716 & 1.175978672140716\\
4 & 1.1757390 & 1.175787 & 1.175978672140716 & 1.175978672140716\\
5 & 1.1750573 & 1.175223 & 1.175406180662251 & 1.175406180662252\\
6 & 1.1750573 & 1.175121 & 1.175406180662251 & 1.175406180662251\\
7 & 1.1743688 & 1.174437 & 1.174833410211082 & 1.174833410211083\\
8 & 1.1743688 & 1.174432 & 1.174833410211082 & 1.174833410211082\\
9 & 1.1743688 & 1.174428 & 1.174833410211082 & 1.174833410211082\\
10 & 1.1736743 & 1.173737 & 1.174260360378985 & 1.174260360378986\\
11 & 1.1736743 & 1.173729 & 1.174260360378984 & 1.174260360378984\\
12 & 1.1736743 & 1.173718 & 1.174260360378983 & 1.174260360378983\\
13 & 1.1729750 & 1.173021 & 1.173687030756756 & 1.173687030756755\\
14 & 1.1729750 & 1.173010 & 1.173687030756751 & 1.173687030756749\\
15 & 1.1729750 & 1.172995 & 1.173687030756742 & 1.173687030756744\\
16 & 1.1729750 & 1.172990 & 1.173687030756736 & 1.173687030756735\\
17 & 1.1722714 & 1.172322 & 1.173113420934324 & 1.173113420934326\\
18 & 1.1722714 & 1.172273 & 1.173113420934253 & 1.173113420934251\\
19 & 1.1722714 & 1.172261 & 1.173113420934170 & 1.173113420934170\\
20 & 1.1722714 & 1.172247 & 1.173113420934124 & 1.173113420934125\\\hline
\end{tabular}
\par
(a) \medskip
\par%
\begin{tabular}
[c]{cc}%
\begin{tabular}
[c]{cll}\hline
Mode & Exact & SPPS\\\hline
36 & 1.171390906148080 & 1.171390906148080\\
37 & 1.170816171958883 & 1.170816171958885\\
38 & 1.170816171841807 & 1.170816171841808\\
39 & 1.170816171670561 & 1.170816171670562\\
40 & 1.170816171522275 & 1.170816171522275\\
41 & 1.170816171436808 & 1.170816171436808\\
42 & not computed & 1.170816171405786\\
43 & 1.170241157428141 & 1.170241157424699\\
44 & 1.170241157130783 & 1.170241157130768\\
45 & 1.170241156386745 & 1.170241156386736\\
46 & 1.170241155534945 & 1.170241155534942\\
47 & 1.170241154879515 & 1.170241154879516\\
48 & 1.170241154528823 & 1.170241154528822\\
49 & not computed & 1.170241154407888\\
50 & 1.169665868832066 & 1.169665868832062\\\hline
\end{tabular}
&
\begin{tabular}
[c]{cll}\hline
$m$ & Exact & SPPS\\\hline
0 & 1.177122819807467 & 1.177122819807467\\
0 & 1.175978672140716 & 1.175978672140716\\
0 & 1.174833410211082 & 1.174833410211083\\
0 & 1.173687030756757 & 1.173687030756755\\
0 & 1.172539530501824 & 1.172539530501818\\
0 & 1.171390906243938 & 1.171390906243931\\
0 & 1.170241157428142 & 1.170241157424699\\
0 & 1.169090334155869 & 1.169090334345329\\
0 & 1.167939131914112 & 1.167939133903106\\
0 & 1.166794286172046 & 1.166794294097307\\
0 & 1.165728772787623 & 1.165728808266585\\\hline
10 & 1.171390906148080 & 1.171390906148080\\
10 & 1.170241154528824 & 1.170241154528824\\
10 & 1.169090277987715 & 1.169090277987714\\
10 & 1.167938398976696 & 1.167938398976695\\
10 & 1.166787247703049 & 1.166787247703050\\
10 & 1.165655359303853 & 1.165655359303857\\\hline
\end{tabular}
\\
(b) & (c)
\end{tabular}
\caption{Mode-propagation constants for the fiber from Subsection
\ref{Example 1} computed using the SPPS method, exact values and the results
from \cite{Liu et al} obtained using WKB and FEM methods. Table (a): first 20
modes; table (b): modes from 36 to 50; table (c): modes corresponding to $m=0$
(``the most inaccurate'') and to $m=10$ (last $m$ for which Wolfram Mathematica
could find the exact characteristic equation). All values must be multiplied
by $10^{7}$m$^{-1}$.}%
\label{Table 1}%
\end{table}%

\begin{figure}
[ptb]
\centering
\includegraphics[
height=3.1996in,
width=5.6704in
]%
{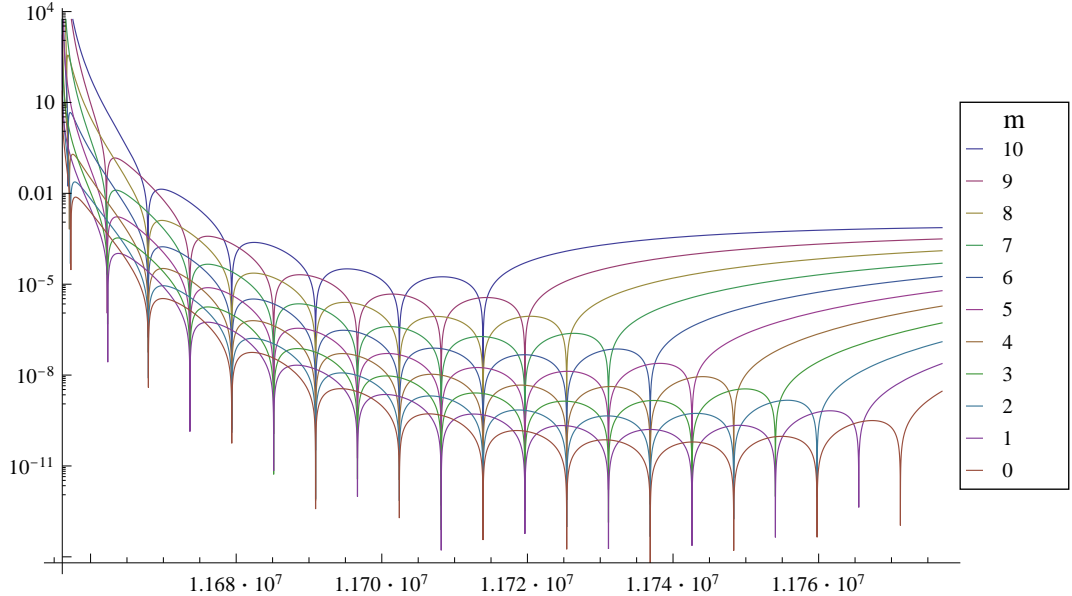}%
\caption{The graphs of the absolute value of the characteristic function for the
problem from Subsection \ref{Example 1} for different values of $m$. }%
\label{Figure 1}%
\end{figure}

Our program has found 121 propagation constants ending with $m=20$, the
overall computation time was about 3 minutes. The number of modes is in a good
agreement with that obtained from the WKB approximation \cite[3.7.2]%
{Okamoto}. Indeed, taking into account that there are 2 conventional modes
corresponding to $m=0$ and 4 corresponding to $m\ge1$ (see \cite[3.4.1]%
{Okamoto}), our computation found 462 conventional modes, while the WKB
approximation \cite[(3.187)]{Okamoto} gives 442 conventional modes. In Table
\ref{Table 1} we show our results together with the exact values and the
values from \cite{Liu et al} obtained with the use of the WKB and the Finite
Element (FEM) methods. The exact values were computed with the help of the
commands \texttt{DSolve} and \texttt{FindRoot} from Wolfram Mathematica 8.0.
It is worth mentioning that Mathematica was only able to compute the
propagation constants for modes up to $m=10$. That explains the missing exact
values for modes 42 and 49 in Table \ref{Table 1}. The outstanding accuracy
achieved by the SPPS method can be appreciated. In Table \ref{Table 1} (c) we
mention that the propagation constants obtained for $m=0$ were the most
inaccurate. Such phenomenon has already been observed in \cite[Example 7.5]{CKT2013}
and the accuracy can be improved by taking more points to represent the formal
powers $\widetilde X^{(j)}$. The accuracy of the obtained propagation
constants for larger values of $m$ was excellent even taking less points
representing the formal powers. It is also remarkable that the propagation
constants conform clusters so that the modes in the same group will propagate
with very similar constants. We illustrate this additionally on Figure
\ref{Figure 1} where the graphs of the absolute value of the characteristic
function \eqref{characteristic equation} are presented for different values of
$m$.

\subsection{Dispersion curves of LP modes}\label{Example 2}

The SPPS method allows us not only to find the propagation constants for a
particular wavelength but also to analyze the behavior of the propagation
constants for a range of wavelengths.

\begin{figure}
[ptb]
\centering
\includegraphics[
height=3in,
width=5in
]%
{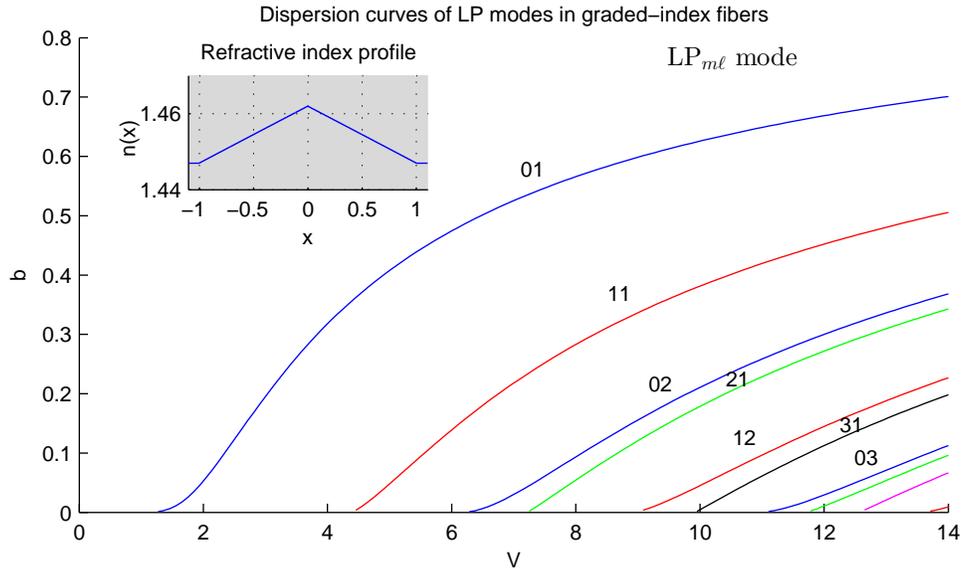}%
\caption{Normalized propagation constant for different propagation modes for
the triangular profile from Subsection \ref{Example 2}.}%
\label{Figure 2}%
\end{figure}

For a numerical illustration we considered the profile \eqref{PowerProfiles}
for $\alpha=1$ (the so-called triangular profile) with $n_{1}=1.462$,
$n_{2}=1.447$, $a=12.5\,\mu$m. On Figure \ref{Figure 2} we present the graphs of the
normalized propagation constant
\[
b=\frac{(\beta^{2}/k^{2})-n_{2}^{2}}{n_{1}^{2}-n_{2}^{2}}%
\]
versus normalized frequency $V=ak\sqrt{n_{1}^{2}-n_{2}^{2}}$ for different
propagation modes. The parameters of the SPPS implementation were $M=10001$,
$N=100$ and we used 140 different values of $V$ to calculate the propagation constants. We refer the reader to \cite{GPM1977} where the normalized cut-off frequency $V\approx 4.381$ of the single-mode operation was computed for the triangular profile fiber. Our results agree with this value.

\subsection{The group velocity and the waveguide dispersion}\label{Example 3}

\begin{figure}
[ptb]
\centering
\includegraphics[
height=3.0364in,
width=5.0401in
]%
{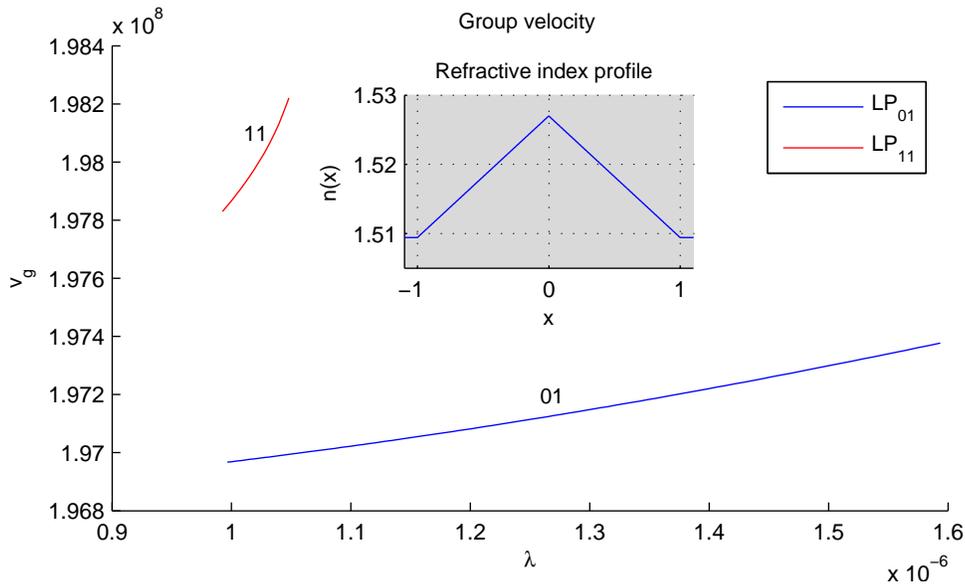}%
\caption{Group velocity for a fiber having a triangular refractive index
profile from Subsection \ref{Example 3}. The refractive index profile
is displayed in the embedded figure.}
\label{GroupVel triangular}%
\end{figure}

\begin{figure}
[ptb]
\centering
\includegraphics[
height=3.0364in,
width=5.0401in
]%
{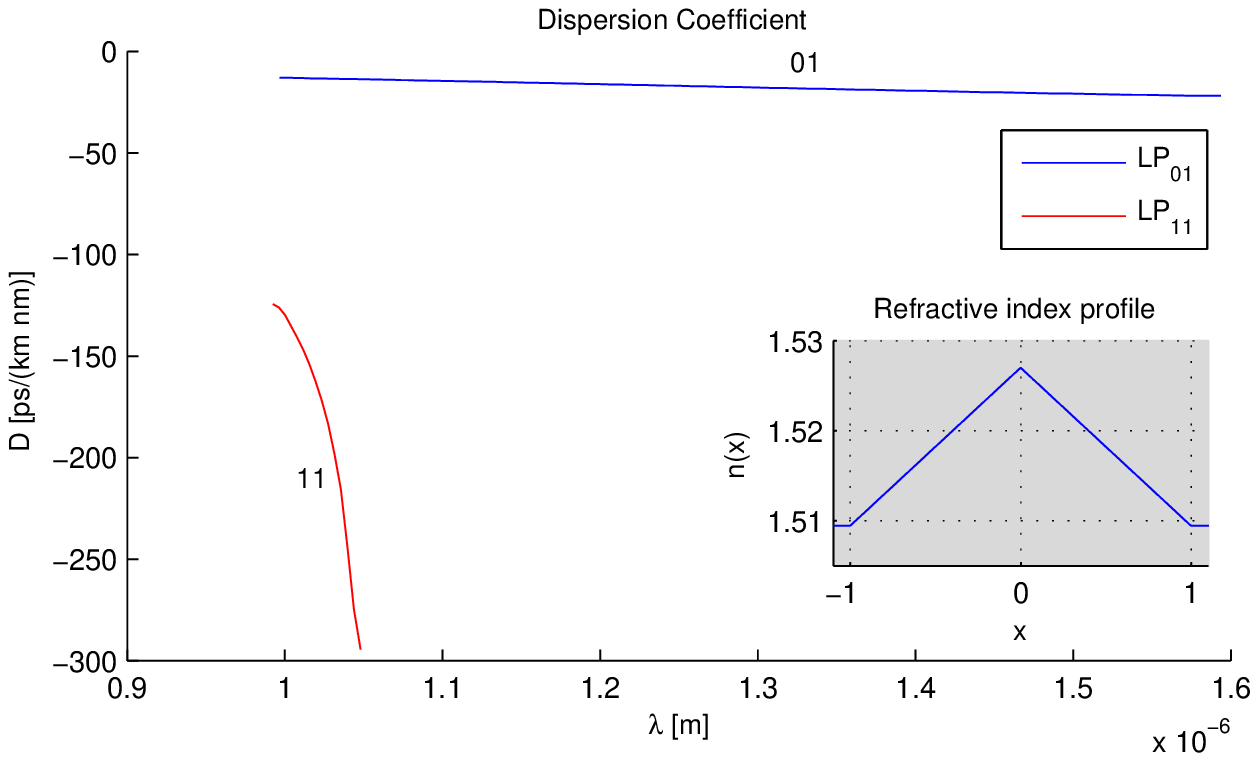}%
\caption{Dispersion coefficient for a fiber having a triangular refractive index
profile from Subsection \ref{Example 3}. The refractive index profile is displayed in the embedded figure.}%
\label{Dispersion triangular}%
\end{figure}

The SPPS method was used to evaluate the group velocity (\ref{Group Velocity}) and the dispersion coefficient (\ref{DispersionParameter}) for the triangular profile (with the refractive index defined by (\ref{PowerProfiles}) with $\alpha=1$).
The parameters of the fiber were taken from \cite{Ainslie Day} and were the following: $n_{1}=1.527,$
$n_{2}=1.5094395,$ $a=3.2\,\mu$m. For the computation we used  $1.00\leq\lambda\leq1.60\,\mu$m, 1000
values for $k$, 100 values for $\beta$, $M=10001$ and $N=100$. The elapsed time was 0.56
minutes. The group velocity is shown on Figure \ref{GroupVel triangular}, and the dispersion parameter is shown on Figure \ref{Dispersion triangular}.
In both graphs the refractive index profile of the fiber was included in a subplot.


%

%

%
%
%
%


\subsection{Simultaneous analysis of material and waveguide dispersion}\label{Example 4}
The refractive index depends on the wavelength. Such dependence is usually described by the Sellmeier equation, see, e.g., \cite[(7.126)]{Adams} or by its generalizations allowing graded-index fibers where dopant concentrations depend on the radius, see \cite{Fleming1984}, \cite{HammondNorman1977}, \cite{Wemple1979}. The SPPS method applied as described in Remark \ref{Remark another transformation} can be used to study the combined material and waveguide dispersion (chromatic dispersion).

\begin{figure}
[ptb]
\centering
\includegraphics[
height=3.0364in,
width=5.0401in
]%
{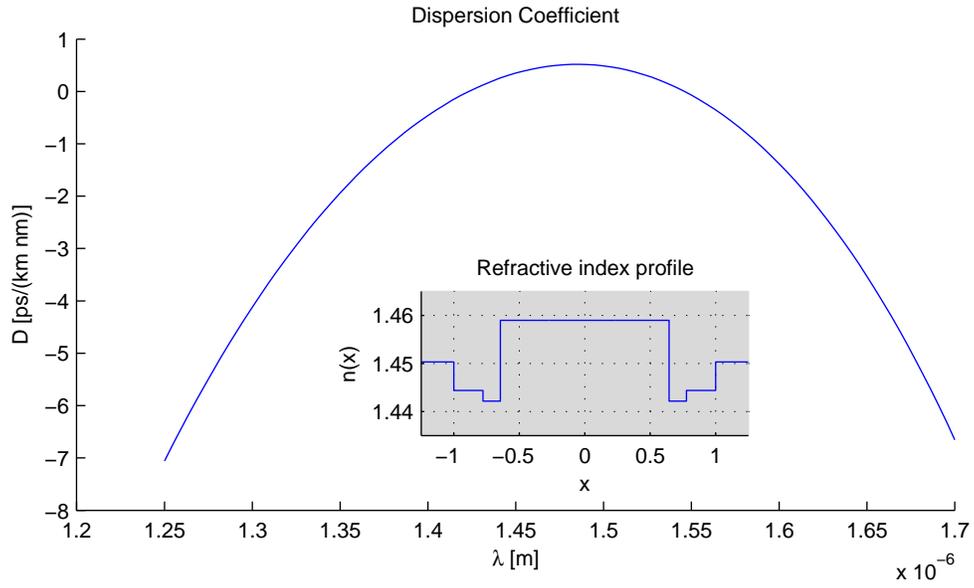}%
\caption{Dispersion coefficient for a triple clad optical fiber. The refractive
index profile is displayed in the embedded figure.}%
\label{Dispersion flattened}%
\end{figure}

\begin{figure}
[ptb]
\centering
\includegraphics[
height=3.0364in,
width=5.0401in
]%
{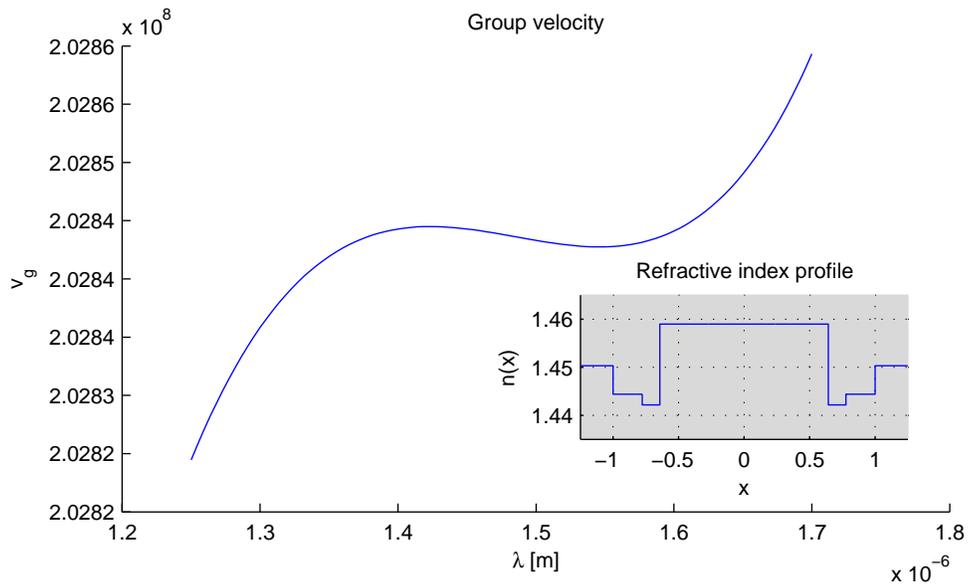}%
\caption{Group velocity for a triple clad optical fiber. The refractive index
profile is displayed in the embedded figure.}%
\label{GroupVel flattened}%
\end{figure}

As an example we considered a dispersion-flattened triple clad fiber (fiber 2 from \cite{Barake1997}). The dopant concentrations of the fiber are the following: core -- 9.1 m/o $P_2O_5$, cladding 1 -- 13.5 m/o $B_2O_3$, cladding 2 -- quenched silica, cladding 3 -- 4.1 m/o $GeO_2$, the radiuses are $r_1=2.9\,\mu$m, $r_2=3.5\,\mu$m and  $r_3=4.5\,\mu$m correspondingly. The Sellmeier coefficients for all mentioned materials were taken from \cite[Table 7.3]{Adams}. For the computation we used $1.25\le \lambda\le 1.7\,\mu$m, 100 values of $k$, 200 values for the search of $\beta$'s, $M=10001$ and $N=100$. The elapsed time was 0.2 minutes.
Dispersion coefficient is shown on Figure \ref{Dispersion flattened}
, and the group velocity is shown on Figure \ref{GroupVel flattened}. In both
graphs the refractive index profile of the fiber was included in a subplot.
As can be seen from the graph, the dispersion is within $\pm 1$ ps/(km$\cdot$nm) for $1.382\le \lambda\le 1.586\,\mu$m.

%

\section{Conclusions}
The SPPS method is developed and applied to problems of wave propagation in graded-index optical fibers. The numerical examples presented show that the method provides a remarkable accuracy achievable within seconds. It shares with the widely used WKB approach the possibility to work with an analytic representation for the solution and for the characteristic function of the problem, offering at the same time the accuracy superior to other available purely numerical methods such as finite element method. We believe that the SPPS approach will become a standard tool for analysis and design of optical fibers and other inhomogeneous cylindrical waveguides.

%

\section*{Acknowledgements}

R. Castillo would like to thank the support of the SIBE and EDI programs of
the IPN as well as that of the project SIP 20140733. Research of V. Kravchenko
and S. Torba was partially supported by CONACYT, Mexico via the projects
166141 and 222478.


\begin{thebibliography}{99}                                                                                               %
\itemsep= 0pt
\small

\bibitem{Adams}M. J. Adams. \emph{An Introduction to Optical Waveguides}, New York: Wiley, 1981.

\bibitem{Agrawal}G. P. Agrawal. \textit{Fiber-optics communication systems},
3rd edition, New York: Wiley, 2002.

\bibitem{Ainslie Day}B. J. Ainslie and C. R. Day. \textit{A review of
single-mode fibers with modified dispersion characteristics}. J.
Lightwave Technol. 4 (1986), no. 8, 967--979.

\bibitem{Barake1997}T. Barake. \emph{A generalized analysis of multiple-clad optical fibers with arbitrary step-index profiles and applications}, MSc. thesis, Virginia Polytechnic Institute, 1997.

\bibitem{Born}M. Born and E. Wolf. \textit{Principles of optics}, 7th ed.,
New York: Cambridge University Press, 1999.

\bibitem{BoumenirChanane}A. Boumenir, B. Chanane. \textit{Computing
eigenvalues of Sturm-Liouville systems of Bessel type}. Proc. Edinburgh Math. Soc. 42 (1999) 257--265.

\bibitem{CKKO2009}R. Castillo-Perez, K. V. Khmelnytskaya, V. V.
Kravchenko and H. Oviedo. \textit{Efficient calculation of the reflectance and
transmittance of finite inhomogeneous layers}. J. Opt. A: Pure and Applied
Optics, 11 (2009), 065707. 


\bibitem{CKT2013}R. Castillo-Perez, V. V. Kravchenko, S. M. Torba.
\emph{Spectral parameter power series for perturbed Bessel equations}, Appl.
Math. Comput. 220 (2013) 676--694.

\bibitem{Chebli1994}H. Ch\'{e}bli, A. Fitouhi and M. M. Hamza.
\textit{Expansion in series of Bessel functions and transmutations for
perturbed Bessel operators}. J. Math. Anal. Appl., 181 (1994), no. 3, 789--802.

\bibitem{DavisRabinovich}P. J. Davis, P. Rabinowitz. \textit{Methods of
numerical integration}, 2nd ed., New York: Dover Publications, 2007.

\bibitem{Fleming1984} J. W. Fleming. \emph{Dispersion in GeO$_2$--SiO$_2$ glasses}. Appl. Opt. 23 (1984), no. 24, 4486--4493.

\bibitem{GPM1977} W. A. Gambling, D. N. Payne, H. Matsumura. \emph{Cut-off frequency in radially inhomogeneous single-mode fibre}. Electron. Lett. 13 (1977) 139--140.

\bibitem{Guillot 1988}J.-C. Guillot, J. V. Ralston. \textit{Inverse spectral
theory for a singular Sturm-Liouville operator on [0,1]}. J. Differ.
Equations 76 (1988), no. 2, 353--373.

\bibitem{HammondNorman1977}C. R. Hammond, S. R.  Norman. \emph{Silica based binary glass systems -- Refractive index behaviour and composition in optical fibres}. Opt. Quantum Electron. 9 (1977) 399--409.

\bibitem{KKB2013}K. V. Khmelnytskaya, V. V. Kravchenko and J. A.
Baldenebro-Obeso. \textit{Spectral parameter power series for fourth-order
Sturm-Liouville problems}, Appl. Math. Comput. 219 (2012) 3610--3624.

\bibitem{KKR2012}K. V. Khmelnytskaya, V. V. Kravchenko and H. C. Rosu.
\textit{Eigenvalue problems, spectral parameter power series, and modern
applications}. Math. Method Appl. Sci. (2014), published online, doi:10.1002/mma.3213.

\bibitem{KosTesh2011}A. Kostenko, G. Teschl. \textit{On the singular
Weyl-Titchmarsh function of perturbed spherical Schr\"{o}dinger operators}, J.
Differ. Equations 250 (2011) 3701--3739.

\bibitem{KrCV08}V. V. Kravchenko. \textit{A representation for solutions of
the Sturm-Liouville equation}, Complex Var. Elliptic Equ. 53
(2008) 775--789.

\bibitem{KrPorter2010}V. V. Kravchenko, R. M. Porter. \textit{Spectral
parameter power series for Sturm-Liouville problems}. Math. Method
Appl. Sci. 33 (2010) 459--468.

\bibitem{Liu et al}Y. E. Liu, B. M. A. Rahman, Y. N. Ning, K. T. V. Grattan.
\textit{Accurate mode characterization of graded-index multimode fibers for
the application of mode-noise analysis}. Appl. Opt. 34 (1995) 1540-1543.

\bibitem{Marcuse}D. Marcuse. \textit{Light transmission optics}. New York:
Van Nostrand Reinhold Company, 1982.

\bibitem{Oh and Paek}K. Oh and U-Ch. Paek. \textit{Silica optical fiber
technology for devices and components}. New Jersey: Wiley, 2012.

\bibitem{Okamoto}K. Okamoto. \textit{Fundamentals of optical waveguides}. San
Diego: Academic Press, 2000.

\bibitem{Weidmann}J. Weidmann. \textit{Spectral theory of ordinary
differential operators}, Lecture Notes in Math., Berlin: Springer, vol. 1258, 1987.

\bibitem{Wemple1979} S. H. Wemple. \emph{Material dispersion in optical fibers}. Appl. Opt. 18 (1979), no. 1, 31--35.

\bibitem{Zettl}A. Zettl. \textit{Sturm-Liouville theory}, Mathematical
Surveys and Monographs, 121. Providence, RI: American Mathematical Society, 2005.
\end{thebibliography}
\end{document}